\documentclass[aps,prl,twocolumn,showpacs,superscriptaddress]{revtex4}

\usepackage{graphicx}

\begin{document}

\title{Slip flow over structured surfaces with entrapped microbubbles}

\author{Jari Hyv\"aluoma}
\affiliation{
Department of Physics, University of Jyv\"askyl\"a, FI-40014 Jyv\"askyl\"a, Finland
}

\author{Jens Harting}
\affiliation{
Institute for Computational Physics, Pfaffenwaldring 27, D-70569 Stuttgart, Germany
}

\date{\today}

\begin{abstract}
On hydrophobic surfaces, roughness may lead to a transition to a
superhydrophobic state, where gas bubbles at the surface can have a strong
impact on a detected slip. We present two-phase lattice Boltzmann
simulations of a Couette flow over structured surfaces with attached gas
bubbles.
Even though the bubbles add slippery surfaces to the channel, they can
cause negative slip to appear due to the increased roughness. The
simulation method used allows the bubbles to deform due to viscous
stresses. We find a decrease of the detected slip with increasing shear
rate which is in contrast to some recent experimental results implicating
that bubble deformation cannot account for these experiments. Possible
applications of bubble surfaces in microfluidic devices are discussed.
\end{abstract}


\maketitle

The no-slip boundary condition states that the fluid velocity at a
fluid-solid interface equals to that of the solid surface.
Although this boundary condition has been 
proven valid at
macroscopic scales, it has no microscopic justification and 
is not fulfilled generally~\cite{Lauga07}. Its validity was debated
already in the early days of fluid mechanics and due to recent
developments in microfluidics the interest in violation of the no-slip
boundary condition has revived~\cite{Neto05}.  In microfluidics, several
experiments have found fluid slip at the boundaries of the flow
channels~\cite{Tretheway02,Joseph06}. As the slip length (defined below)
has typically a magnitude measured in nano- or micrometers, the appearance
of slip does not have noticeable ramifications in macroscopic flows.
However, in microfluidic devices with large surface-to-volume ratio,
surface properties may dramatically affect the flow resistance. The
possibility to engineer 
slip properties in a controlled way is therefore crucial for microfluidic
applications.
 
The slip at fluid-solid boundaries can be quantified by Navier's boundary
condition, which states that the slip velocity is proportional to the
velocity gradient, i.e., $v_s = b (\partial v / \partial z)$ at the
surface $z=z_0$. Here, $b$ is the slip length which is the distance below
the surface where the velocity linearly extrapolates to zero. Higher slip
length means larger slip and lower friction at the boundary.

Mechanisms behind boundary slip include surface roughness and
structural characteristics of roughness~\cite{Kunert07,Ybert07},
roughness-induced dewetting on hydrophobic
surfaces~\cite{Cottin03,Sbragaglia06}, dissolved gas and bubbles on the
surface~\cite{deGennes02,Steinberger07,Lauga04}, as well as wetting
properties~\cite{Harting06}. Usually, roughness decreases the slip
length due to increased dissipation and the roughness induced slip is
just an artifact~\cite{Kunert07}. Theoretically, it was shown that even
slippery surfaces, if rough enough, can provide no-slip
boundaries~\cite{Richardson73}.
However, if surfaces are hydrophobic, roughness may increase the slip due
to a transition to a superhydrophobic (fakir or Cassie)
state~\cite{Lafuma03}.  Here, liquid cannot enter between roughness
elements but stays at the top of them. Thus, gas bubbles or layers are
formed thereby lubricating the flow due to a reduced liquid-solid contact
area. By using a surface patterned with a square array of cylindrical
holes, Steinberger et al. found that gas bubbles may also cause an
opposite effect, i.e., slip is reduced if microbubbles are present in the
holes~\cite{Steinberger07}. Numerically, they found even negative slip lengths for flow over
such a bubble mattress. Negative slip means that the effective no-slip
plane is inside the channel, i.e., the bubbles increase the flow
resistance.  These mechanisms are related to so-called effective slip and
should be distinguished from the (smaller) intrinsic slip on smooth
surfaces.
Another peculiarity in the boundary slip is the shear-rate dependence
observed in some experiments but not in the others~\cite{Neto05}. The question
if the shear-rate dependence is a true property of slip is still to be answered.

As the different mechanisms behind the slip phenomenon are strongly
intertwined, the experimental study of a single mechanism is a complicated
task. Therefore, numerical simulations are attractive as they
provide a controllable way to change a single property of fluid or surface
while keeping the others unchanged. Most computer simulations so far have
been performed using molecular
dynamics~\cite{Koplik95,Thompson97,Barrat99,Cottin03}. For computational
reasons molecular dynamics is limited to length scales of tens of
nanometers and time scales of nanoseconds, which do not comply with the
experimentally relevant scales. Therefore, mesoscopic lattice Boltzmann
(LB) simulations have recently been applied to study flow in microchannels or
along hydrophobic
surfaces~\cite{Kunert07,Sbragaglia06,Harting06,Benzi06,Hyvaluoma07}.  This
method allows to reach experimentally relevant scales and preserves those
interactions needed to describe the underlying physics.

Our simulations utilize the multiphase LB model by Shan and Chen~\cite{Shan94}. 
Dynamics of the method is governed by a discretized Boltzmann equation
\begin{equation}
\label{eqlbgk}
f_i({\bf r} + {\bf c}_i,t+1) - f_i({\bf r},t)
= - \frac{1}{\tau} \left[ f_i({\bf r},t) - f_i^{eq}({\bf r},t)
\right],
\end{equation}
where $f_i$ is a 
distribution function describing the probability to find a particle at position 
${\bf r}$ at time step $t$, moving in lattice direction ${\bf c}_i$. We use a 
three-dimensional lattice with 19 discrete velocities. The right-hand side of 
Eq.~\ref{eqlbgk} models the relaxation of the $f_i$ towards a local equilibrium 
due to collisions among the particles on a time scale given by the relaxation 
time $\tau$. 
The mean-field interactions between particles are given by a force
\begin{equation}
{\bf F} = G_b \psi({\bf r}) \sum_i t_i
          \psi({\bf r} + {\bf c}_i) {\bf c}_i,
\end{equation}
where $\psi = 1 -\exp{(-\rho/\rho_0)}$ is an effective mass ($\rho$ is
the fluid density, $\rho_0$ a reference density), 
$G_b$ tunes
the strength of the interaction, and $t_i$'s are weight factors for
different lattice directions. This force term leads to a non-ideal
equation of state with pressure $P = c_s^2\rho +
\frac{1}{2}c_s^2G_b\psi^2$, where $c_s$ is the speed of sound,
and it enables simulations of liquid-vapor systems with
surface tension. To model the wetting behavior at
fluid-solid surfaces, a similar interaction is added between the fluid
and solid phases, and the contact angle is tuned by setting a
density value $\rho_w$ at the boundaries~\cite{Benzi06b}.

In this article we investigate liquid slip in Couette flow, where the flow
is confined between two parallel walls. One of the walls is patterned with
holes and vapor bubbles are trapped to these holes. Steinberger et
al.~\cite{Steinberger07} presented finite-element simulations of flow over
rigid ``bubbles'' by applying slip boundaries at static bubble surfaces.
The LB method allows the bubbles to deform if the viscous forces are high
enough compared to the surface tension. We are also interested in how
surface patterning affects the slip properties of these surfaces, and how
bubbles could be utilized to develop surfaces with special properties for
microfluidic applications.

\begin{figure}
\includegraphics[width = 0.35\textwidth]{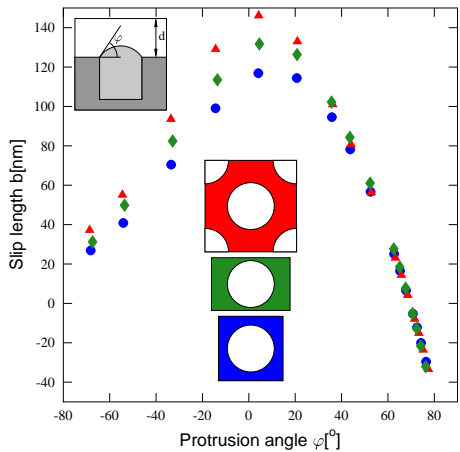}
\caption{\label{fig:slip}
(Color online) Slip length $b$ as a function of protrusion angle
$\varphi$.
A unit cell of each array is shown in insets and corresponding
results are given by triangles (rhombic array), diamonds (rectangular
array), and circles (square array). The inset in the top-left corner
shows the definition of $\varphi$.}
\end{figure}

In our simulations, the lower wall is static and has the
topographical patterning whereas the upper one is smooth and moved
with velocity $u_0$.
The distance between walls is $d=1$~$\mu$m (40 nodes) in all simulations,
and the area fraction of holes $0.43$ unless stated otherwise. The system
boundaries are periodic and a unit cell of the regular array is included
in a simulation. To trap bubbles to holes, some heterogeneity is needed at
the edges of holes in order to pin the contact line. To this end, we use
different wettabilities for boundaries in contact with the main channel
and with the hole. The protrusion angle $\varphi$ (see Fig.~\ref{fig:slip}
for definition) is varied by changing the bulk pressure of the liquid. A
similar technique to form bubbles on structured surfaces was used
experimentally by Bremond et al.~\cite{Bremond06}. The effective slip
length can be calculated from the shear stress $\sigma = \mu {\rm d}v/{\rm
d}z$ acting on the upper wall, which is obtained from the no-slip boundary
condition imposed at the fluid-solid boundaries. Thus the effective slip
length reads as $b = \mu u_0/ \sigma - d$, where $\mu$ is the dynamic
viscosity of the liquid.
$b$ is measured from the top of the structured surface and $G_b$ is
chosen such that the density ratio between liquid and gas is
$22$. This ratio is too small for a realistic description of gas bubbles
in a liquid. Also, the interface between both phases is of finite width
causing the resistance in the vapor phase.
These limitations of multiphase LB models do not
influence the qualitative insight obtained from our simulations. 

\begin{figure*}
\includegraphics[width = 0.95\textwidth]{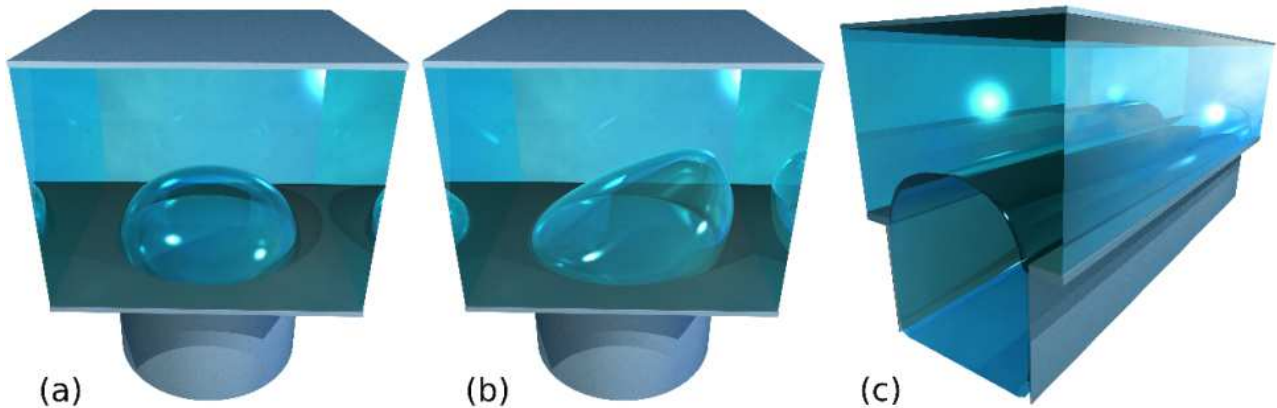}
\caption{\label{fig:vis}
(Color online)
Snapshots of simulations of bubbles on structured surfaces.  Shown are a
square array of bubbles with Capillary number (a) $Ca = 0.02$ and (b) $Ca
= 0.4$, and (c) a slot with a cylindrical bubble. In each case a unit cell
is shown.
}
\end{figure*}

In order to study the effect of a modified protrusion angle and different
surface patternings, we use three different arrays of bubbles, i.e., a
square array, a rectangular array where the distance of bubbles is larger
in one direction than in the other, and a rhombic array. These surfaces
have cylindrical holes with radius $a = 500$~nm and the area fraction of
the holes is equal in all cases. The shear rate is such that the Capillary
number $Ca = 0.16$. The Capillary number is the ratio of viscous and
surface forces, i.e., $Ca = \mu a G_s /\gamma$, where $G_s$ and $\gamma$
are the shear rate and surface tension, respectively.  Snapshots of
simulations are shown in Fig.~\ref{fig:vis} and the slip lengths obtained
are shown in Fig.~\ref{fig:slip}. The observed behavior is similar to that
reported in~\cite{Steinberger07}, where a square array of holes was
studied. In particular, we observe that when the protrusion angle is large
enough the slip length becomes negative. We also see that the maximum of
the slip length is obtained when the protrusion angle equals zero.  Since
the area fraction of the bubbles is the same in all three cases, the
results clearly indicate that slip properties of the surface can be
tailored not only by changing the protrusion angle but also by the array
geometry. 

\begin{figure}
\includegraphics[width = 0.43\textwidth]{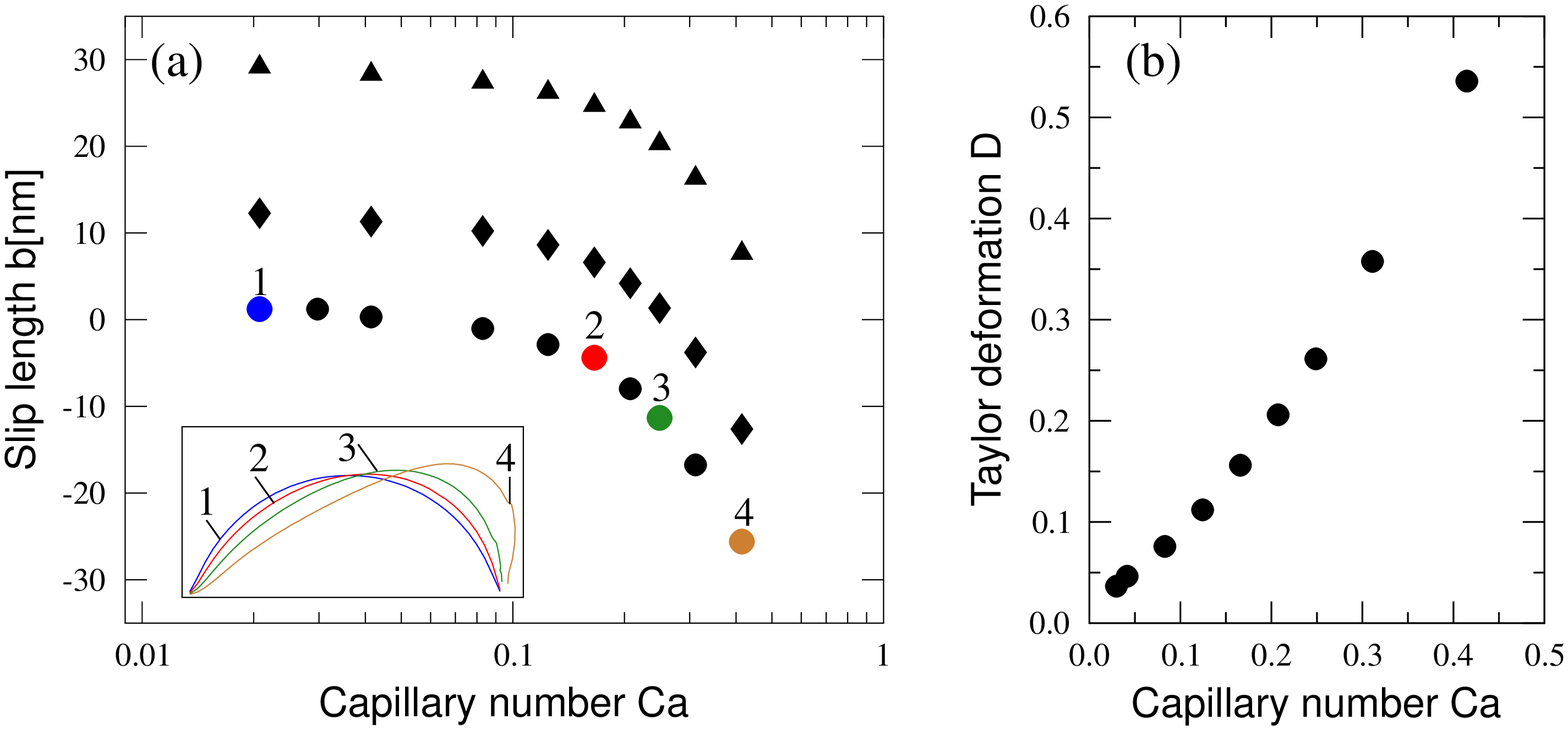}
\caption{\label{fig:shear}
(Color online) (a) Slip length as a function of capillary 
number for a square array of bubbles with three different 
protrusion angles, $\varphi = 63^\circ, 68^\circ,$ and 
$71^\circ$ (from uppermost to lowermost).
In the inset shown are cross sections of liquid-gas interfaces
for four capillary numbers.
(b) Taylor deformation as a function of Capillary number
for the bubble with $\varphi = 71^\circ$.
These Ca values correspond to shear rates of the order of 
$10^{-6} - 10^{-7}$ s$^{-1}$.
}
\end{figure}

Next, the shear-rate dependence of the slip length is investigated.  As
the shear rate and thus the viscous stresses grow the bubbles are deformed
(see Fig.~\ref{fig:vis}) and the flow field is modified. 
Results are shown in
Fig.~\ref{fig:shear}a, where the slip length is given as a function of the
Capillary number for three different protrusion angles. We also calculate
the Taylor deformation $D = (\ell-a)/(\ell+a)$ of the bubbles by fitting
an ellipse to the bubble interface. Here,
$a$ and $\ell$ are the minor and major axes of the ellipse,
respectively. A slightly superlinear relation between $D$ and $Ca$ is
observed (Fig.~\ref{fig:shear}b) in accordance with a two-dimensional
case in Ref.~\cite{Feng94}. 
We find that increasing shear rates cause the slip length to decrease.
This behavior is contradictory to those found
in some experiments using surface force apparatuses (see, e.g.,
Ref.~\cite{Zhu01}), where a strong increase in the slip is observed after
some critical shear rate. This shear-rate dependence has been explained,
e.g., with formation and growth of bubbles~\cite{deGennes02,Lauga04}.
However, one should notice that these experiments are dynamic in nature
while we simulate a steady case. In our simulations there is no formation
or growth of the bubbles but we determine the slip for given bubbles of
given size. Our results indicate that the deformation of the bubbles and
the changes in the flow field thereby occurring cannot be an explanation
for the observed shear-rate dependence.
On the other hand, our results are consistent with~\cite{Kunert07}, where 
it is shown that smaller roughness leads to smaller values of a detected
slip. In the present case, the shear reduces the average height of the bubbles and thus
the average scale of the roughness decreases as well. 

\begin{figure}
\centerline{\includegraphics[width = 0.36\textwidth]{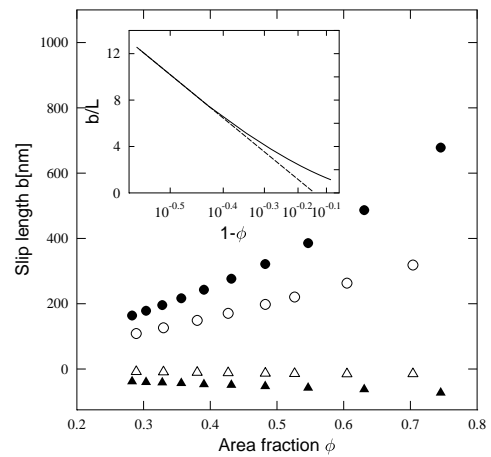}}
\caption{\label{fig:slot}
Slip length as a function of hole area fraction.  Filled circles denote
values for the flow parallel to the slots and triangles relate to the
perpendicular direction. Corresponding open symbols are for a surface with
rectangular holes ($\varphi = 72^\circ$).  The inset shows the normalized
slip length as a function of the solid-area fraction, where the dashed
line is the theoretical prediction for small solid-area fractions.
}
\end{figure}

As seen above, the slip properties of a bubble mattress can be tailored by
changing the surface patterning. Next, we consider a surface which has
slots with a width of 1~$\mu$m, and cylindrical bubbles protruding to the
channel with an angle $\varphi = 72^\circ$ (Fig.~\ref{fig:vis}c). We
apply shear parallel and perpendicular to the slots, while the area fraction of
the bubbles ($\phi$) is varied by changing the distance between the grooves ($L$). 
According to our results (Fig.~\ref{fig:slot}), $b$ strongly depends on
the flow direction and even the qualitative behavior changes. When the
flow is parallel to the slot a positive slip is observed, but for
perpendicular flow the slip becomes negative. This observation can be
explained by means of theoretical predictions of Richardson, who showed
that slippery surfaces lead to (macroscopic) no-slip boundaries if rough
enough~\cite{Richardson73}. Our results support these predictions. In the
case with flow parallel to the bubbles the streamlines are straight and
the flow does not ``see'' any roughness. However, in the perpendicular
direction roughness is caused by the bubbles and negative slip is
observed. The inset of Fig.~\ref{fig:slot} depicts the scaling behavior
of the slip length as a function of the solid-area fraction for flow
perpendicular to the slots. 
We see that for small solid-area
fraction the slip length obeys the scaling law $b \sim -L \log (1-\phi)$
as was recently predicted in Ref.~\cite{Ybert07}.  Anisotropic behavior
has been observed in the case of flat surfaces composed of stripes with
no-slip and perfect-slip boundary
conditions~\cite{Philip72,Lauga03}. Our simulations differ from these
studies as our bubbles protrude to the channel thus leading to a larger
effect than for a flat surface.  In order to understand less idealized
surface patternings, we study flow over rectangular holes. The aspect
ratio of the holes is chosen such that the longer side is three times the
smaller one. We observe similar qualitative behavior (Fig.~\ref{fig:slot}), 
but the difference in slip length between the two
flow directions is less pronounced. Obviously, by changing the aspect
ratio of the hole, the anisotropic behavior of the slip can be tailored.
Due to the striking difference between the slip properties in the two
perpendicular directions, we believe that this kind of
surfaces may find applications in microfluidic devices. Anisotropic
surfaces could be used to suppress the flow in an unwanted direction while
enhancing it in another one. The suppressing behavior is further amplified
by the shear-rate dependence, since the negative slip is growing with the
shear rate.

To conclude, we simulated Couette flow in a microchannel where one of the
walls is patterned and microbubbles are attached to the pattern. We found
that the slip properties of the surface can be tailored by changing the
hole array and that such a surface with bubbles may yield negative slip,
i.e., increased resistance to flow, if bubbles are strongly protruding to
the channel.  Our results can be qualitatively compared to previous
results~\cite{Steinberger07}, but overcome their limitation of a static
liquid-vapor interface. This allowed to study the influence of the shear
rate on the deformation of the interface and it's effect on the measured
slip. We found that the slip decreases with increasing shear rate
demonstrating that shear induced bubble deformation cannot explain recent
experimental findings where slip increases with increasing shear
rate~\cite{Zhu01}.
We proposed a special surface patterning which can be used to produce
surfaces where the slip is positive in one direction and negative in the
perpendicular one. Such a surface might be useful to construct
microfluidic devices with tunable flow throughput which could be
controlled by adding bubble ridges parallel or perpendicular to the flow.
In addition, we have shown that the throughput could be tailored by tuning
the bulk pressure, i.e., the protrusion angle, or the shear rate.

\begin{acknowledgments}
We are grateful to Jyrki Hokkanen (CSC -- Scientific Computing Ltd.,
Espoo, Finland) for the bubble visualizations.  This work was supported by
the Academy of Finland (Project No.  7117283), the collaborative research
center 716 and the DFG program ``nano- and microfluidics''. First author
thanks the staff of ICP, Stuttgart for kind hospitality during his stay.
\end{acknowledgments}

\end{document}